\documentclass{IEEEtran}

\usepackage[table]{xcolor}
\usepackage{cite}
\usepackage{graphicx}
\usepackage{lipsum}
\usepackage{stfloats}
\usepackage{amssymb}
\usepackage{pifont}
\usepackage{bbding}
\usepackage{booktabs}
\usepackage{ctable}
\usepackage{threeparttable}
\usepackage{footmisc}
\usepackage{framed} 
\usepackage{array}
\usepackage{multirow}
\usepackage{longtable}
\usepackage{rotating}
\usepackage{fancyhdr}
\usepackage{lastpage}
\usepackage{layout}
\usepackage{wrapfig}
\usepackage{xcolor}
\usepackage{tabularx}
\usepackage{tabu}
\usepackage{enumitem}
\usepackage{url}
\graphicspath{{Figures/}}
\DeclareGraphicsExtensions{.png, .eps}

\begin{document}
	\title{WiserVR: Semantic Communication Enabled Wireless Virtual Reality Delivery}
	\author{
	Le Xia, Yao Sun, Chengsi Liang, Daquan Feng, Runze Cheng, Yang Yang, and Muhammad Ali Imran
	\thanks{
	
	Le Xia, Yao Sun (\textit{corresponding author}), Chengsi Liang, Runze Cheng, and Muhammad Ali Imran are with University of Glasgow, United Kingdom;
	
	Daquan Feng is with Shenzhen University, China;
	
	Yang Yang is with Beijing University of Posts and Telecommunications, China.
	}}	
	\maketitle
	\begin{abstract}
	Virtual reality (VR) over wireless is expected to be one of the killer applications in next-generation communication networks.
	Nevertheless, the huge data volume along with stringent requirements on latency and reliability under limited bandwidth resources makes untethered wireless VR delivery increasingly challenging.
	Such bottlenecks, therefore, motivate this work to seek the potential of using semantic communication, a new paradigm that promises to significantly ease the resource pressure, for efficient VR delivery.
	To this end, we propose a novel framework, namely WIreless SEmantic deliveRy for VR (WiserVR), for delivering consecutive 360$^{\circ}$ video frames to VR users.
	Specifically, multiple deep learning-based modules are well-devised for the transceiver in WiserVR to realize high-performance feature extraction and semantic recovery.
	Among them, we dedicatedly develop a concept of semantic location graph and leverage the joint-semantic-channel-coding method with knowledge sharing to not only substantially reduce communication latency, but also to guarantee adequate transmission reliability and resilience under various channel states.
	Moreover, implementation of WiserVR is presented, followed by corresponding initial simulations for performance evaluation compared with benchmarks.
	Finally, we discuss several open issues and offer feasible solutions to unlock the full potential of WiserVR.
	\end{abstract}
	
	\section{Introduction}
	Virtual reality (VR) over wireless has become an extremely prevailing application in communication networks that can enrich our intelligent life by empowering users with a panoramic experience, whose market value is expected to reach 87.97 billion dollars by 2025~\cite{guo2021power}.
	However, achieving fully-immersive VR requires the delivery of massive data (in gigabytes) and an ultra-low latency (20ms or less)~\cite{sun2019communications}, thus placing stringent requirements on ultra-reliable and low-latency communications (URLLC)~\cite{she2021tutorial}.
	Further notice a fact that data traffic burden and latency bottleneck mainly originate from the wireless transmission between remote servers and VR users~\cite{elbamby2018toward}.
	Therefore, the irreconcilable contradiction between inherent bandwidth limitations and the exponential growth in demand for VR services is foreseeably exacerbated to the peak in future wireless VR delivery networks.
	
	Fortunately, semantic communication (SemCom), as an emerging paradigm beyond Shannon's classical information theory~\cite{weaver1953recent}, promises to significantly alleviate the scarcity of communication resources~\cite{xia2022wireless,weng2021semantic,huang2021deep}.
	Different from traditional bit-oriented communication~\cite{cai2006efficient}, the focus of SemCom is on the correct reception of implicit meaning in delivered source messages, where efficient semantic coding models (including semantic encoder at the source and semantic decoder at the destination) under equivalent background knowledge will be of critical importance to pursue adequate ``semantic fidelity' and eliminate semantic ambiguity.
	Moreover, through advanced and sophisticated deep learning (DL) algorithms-powered semantic models, the desired meaning delivered via SemCom can still be accurately interpreted even with intolerable bit errors, which errors, however, often cause meaning confusion in the traditional communication.
	
	Inspired by these, SemCom is specifically considered in this work as a viable solution to tackle with the aforementioned dilemmas in traditional VR delivery.
	In a nutshell, the intrinsic reason for this idea lies in the enormous benefits of wireless bandwidth saving and reliability promotion yielded from their combination, making it a promising candidate capable of fulfilling the next-generation URLLC.
	Nevertheless, such a conceptual evolution of perfectly incorporating SemCom with wireless VR delivery still faces several inevitable hurdles, which can be briefly attributed to the following three aspects.
	
	\textbf{Semantic Representation for Transmitted Consecutive Video Frames:}
	Static and dynamic objects may coexist in VR video frames, where the semantics implicit in each static object between different frames are normally identical and the change process of each dynamic object in consecutive frames should be regular.
	Envision that if the semantics of static and dynamic objects are transmitted separately, then considerable bits and transmission latency can be saved.
	Hence, how to devise a relevant semantic encoding model to realize semantic representation for consecutive frames is the first nontrivial point.
		
	\textbf{Channel State-Aware VR Semantic Delivery:}
	Generally, different wireless channels can incur signal attenuation and distortion to differing extents, which may lead to severe semantic ambiguity of received video frames and further greatly affect the final rendered VR quality.
	In order to enable sufficient reliability and robustness, the second challenge is located at how to take account for different channel states in the design of VR semantic delivery.
		
	\textbf{Video Frame Reconstruction from Received Semantic Features:}
	To serve VR users with the optimal quality of experience (QoE), the ability to restore original video frames with high accuracy needs to be guaranteed in the semantic reconstruction model, which requires a key prerequisite of equivalent background knowledge~\cite{xia2022wireless}.
	Additionally, the extra computing latency introduced by semantic models should at least be covered by the reduced communication latency budget, thus forming the third tricky problem.
	
	To the best of our knowledge, no article has explored the potential of using SemCom for VR applications.
	In full view of the characteristics of 360$^{\circ}$ VR video delivery, this article proposes a novel framework, namely WIreless SEmantic deliveRy for VR (WiserVR), with the awareness of URLLC.
	Specifically, the WiserVR concentrates upon a common VR delivery scenario, where consecutive video frames of 2-dimensional (2D) field of views (FOVs) are delivered from the mobile edge computing (MEC) server to VR users.
	By employing multiple state-of-the-art DL algorithms, significant transmission traffic and communication latency can be saved by WiserVR, while ensuring adequate semantic fidelity during VR delivery.
	In short, our main contributions are summarized as follows.
	\begin{itemize}
		\item To achieve high-performance semantic extraction and recovery, a novel transceiver is dedicatedly designed with multiple differing semantic modules.
		Specially, a critical concept of semantic location graph (SLG) is developed to separate the delivery of static and dynamic objects for each 2D FOV tile to provide accurate semantic calibration for video frame reconstruction.
		\item We then demonstrate a viable implementation of WiserVR, which consists of three successive stages of WiseVR Initial Stage, VR Service Preparation Stage, and VR Delivery Stage.
		Moreover, numerical results validate that our WiserVR can save a significant communication delay while maintaining high-precision VR video delivery compared with two benchmarks.
		\item Finally, several open issues with prospects are outlined in terms of semantic metrics, model storage, coding under time-varying channels, and knowledge sharing efficiency.
	\end{itemize}
	
	For the remainder of this article, the process of typical 360$^{\circ}$ VR service provisioning is first presented with its current limitations.
	Then, we shed light on detailed structure of WiserVR, and analyze the potential benefits it can yield in line with URLLC.
	Afterward, a viable implementation of WiserVR is exhibited.
	Finally, we open the doors for research issues with outlook discussed, and close this article with conclusions.
	
	\section{How Wireless VR Delivery Enables SemCom?}
	\subsection{Process of Typical 360$^{\circ}$ VR Service Provisioning}
	In order to bring users a realistic immersive VR experience, a 360$^{\circ}$ VR video needs to go through a series of processing from construction to display.
	The typical process includes: 1) Arranging a multi-camera array to capture the video scene in a real environment; 2) Stitching captured scene images to obtain a spherical video; 3) Performing equirectangular projection to the spherical video to obtain a 2D video, and then evenly dividing it into fixed-duration video segments in time domain, while evenly dividing the segment in each duration into multiple tiles in spatial domain; 4) Extracting all 2D FOVs with respect to given viewpoints from the 2D video, where all tiles covered by each FOV chunk can be pre-cached at the MEC server; 5) Stitching all associated tiles to form an intact 2D FOV, projecting the 2D FOV into a 3D FOV, and finally rendering it to display on VR devices~\cite{sun2019communications}.
    
	Undoubtedly, wireless VR delivery is latency-sensitive with high bandwidth requirements.
	Although various advanced technologies, such as edge computing, device-to-device caching, and high-frequency communication, have played an effective role in reducing bandwidth consumption and transmission latency~\cite{elbamby2018toward}, in the face of massive and dynamic VR requests, several key limitations of traditional delivery schemes still remain, which are summarized as follows.
	
	\textbf{Inevitable Latency Burden:}
	To prevent dizziness, human eyes need to perceive accurate and smooth movements from VR videos with low motion-to-photon latency, thus imposing stringent requirements on both communication and computing latency.
	Besides, notice a fact that high data volume should be the main burden on communication delay.
	Further combined with limited computing capabilities of current VR devices, therefore, the most effective way is to reduce the transmission traffic as much as possible to alleviate heavy latency burden.
	
	\textbf{Ever-Increasing VR Traffic:}
	Although the current FOV-based VR delivery approach significantly reduces transmitted data compared to traditional whole-video delivery, once there is excessive amount of VR users requesting delivery at the same time, the network still faces a tremendous transmission load.
	Meanwhile, video quality at 8 K and above is essential to provide users with higher-resolution VR viewing, for which, however, the data rate needs to be greater than 1 Gbps~\cite{mangiante2017vr}.
	Hence, the endless increase of VR traffic is foreseeable.
	
	\textbf{Insufficient Transmission Reliability:}
	As VR users move freely over the period of use, the network environment surrounding them will also change, thereby their wireless VR delivery may encounter differing physical channels as well as channel states.
	Considering such channel variances, how to guarantee highly reliable and robust VR video transmission in the dynamic environment is another concern.
	
	\subsection{WiserVR Framework}
	
	\begin{figure*}[ht]
		\centering
		\includegraphics[width=1\textwidth]{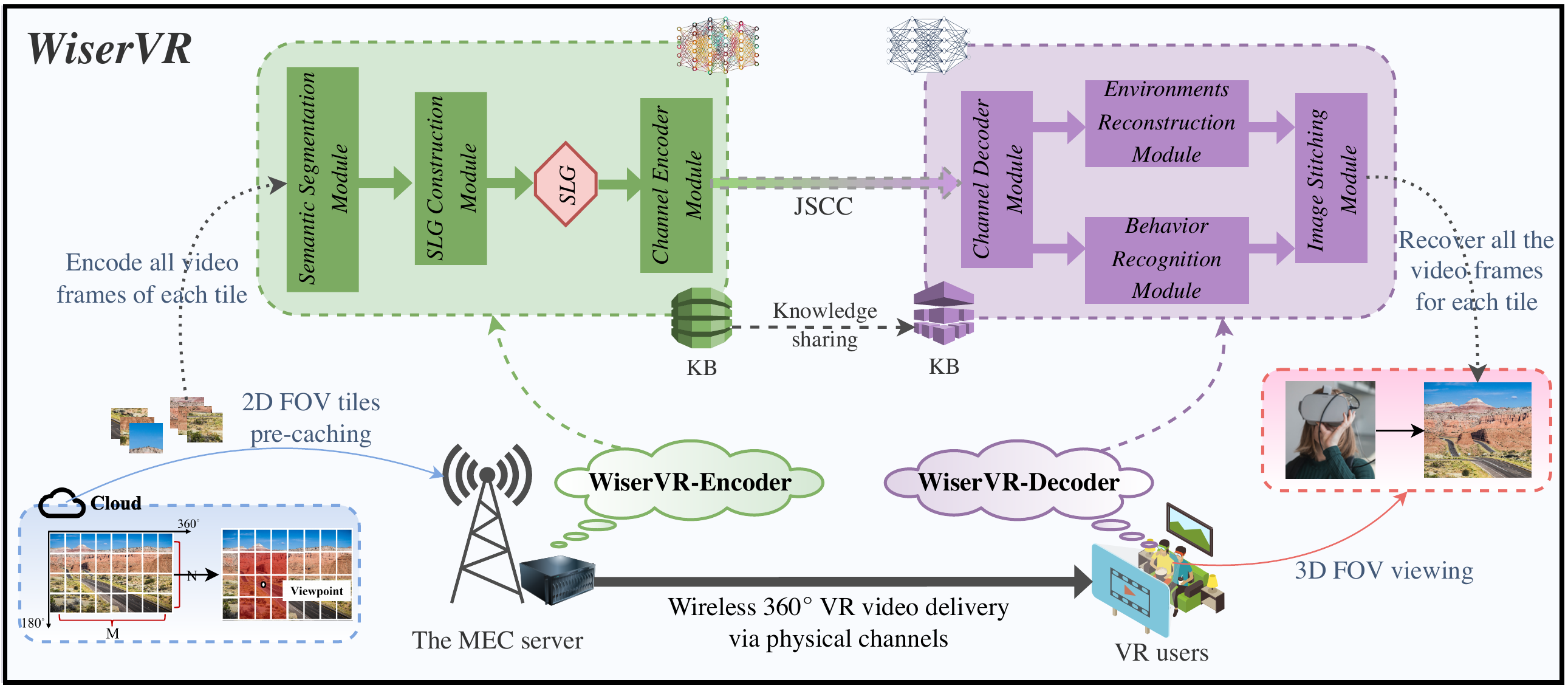} 
		\caption{Overview of the proposed WiserVR framework for delivering 360$^{\circ}$ video streaming from the MEC server to VR users.}
		\label{WiserVR}
    \end{figure*}
	
	Recently, DL-based SemCom has been studied in multiple communication service scenarios, including text~\cite{xia2022wireless}, speech~\cite{weng2021semantic}, image~\cite{huang2021deep}, and so on.
	In a nutshell, the rationale behind the growing prosperity of SemCom lies in the increasingly powerful representation and generalization abilities of DL models.
	Nevertheless, research vacancies for VR in conjunction with SemCom remain and are evidently valuable, especially considering the peculiarity and intractability of semantic representation for real-time video streaming.
	
	In this article, we propose here a novel WiserVR framework as a potentially feasible solution dedicated to 2D FOV video frame delivery.
	As elucidated in Fig.~\ref{WiserVR}, the entire WiserVR encompasses two components: a WiserVR-Encoder network placed at the MEC server and a WiserVR-Decoder network equipped on VR devices.
	In the WiserVR-Encoder, there are three successive modules of semantic segmentation, SLG construction, and channel encoder, each has a differing semantic processing function driven by different DL models.
	Elaborations on these modules and their functions will be provided in the next section.
	It is worth noting here that the SLG is uniquely proposed, which is a multi-node graph with semantic labels attached to each node, and these nodes are generated based on semantic segmentation to all objects in each video frame.
	On this basis, all static and dynamic objects within one tile can be split under the comparison between multiple consecutive frames to be encoded and delivered separately, processing details of which will be given later.
	
	At the VR user side, a corresponding WiserVR-Decoder is devised to recover all the original 2D FOV tiles from received bits.
	This has to go through another four modules of channel decoder, environment reconstruction, behavior recognition, and image stitching, which details will also be introduced later.
	Among them, the WiserVR-Decoder takes full advantage of the semantic calibration function of SLG to help determine the exact location and status of each object in the tile, thereby further enhancing the reliability of VR delivery.
	
	Notably, both WiserVR-Encoder and WiserVR-Decoder are fully pre-trained under their respective knowledge bases (KBs), where each KB can be regarded as an entity storing colossal training data with respect to a particular 360$^{\circ}$ VR video.
	According to the aforementioned knowledge equivalence principle in SemCom, the knowledge sharing process becomes indispensable in the WiserVR framework, which guarantees the highly accurate VR semantic delivery and recovery.
	Furthermore, a joint semantic-channel coding (JSCC) design is applied here~\cite{zhang2022toward}, which can offer WiserVR not only a robust semantic preservation ability, but also a resilience adaptability under different channel states.
	Most importantly, as part of developments for next-generation wireless VR delivery systems, JSCC is capable of well integrating these DL-based semantic coding models with channel encoding and decoding models in both transmitter and receiver to enable wireless SemCom between the MEC server and VR users.
	
	\begin{figure*}[ht]
		\centering
		\includegraphics[width=1\textwidth]{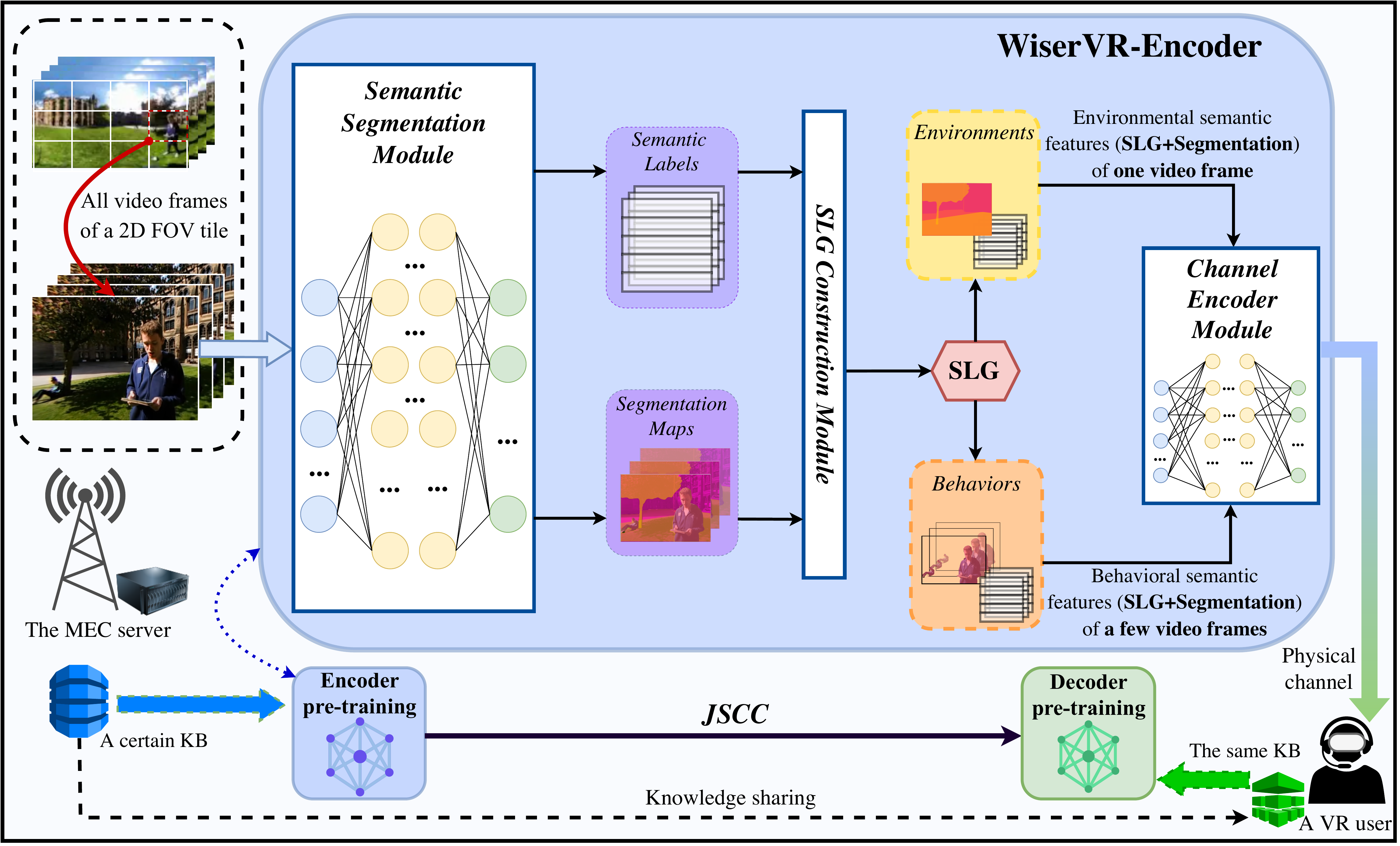} 
		\caption{The detailed structure of the WiserVR-Encoder network.}
		\label{Encoder}
    \end{figure*}
	
	\section{SLG-based Transceiver Design in WiserVR}
	Consider a tile-based 360$^{\circ}$ VR video delivery scenario, where the MEC server acts as the video provider to serve VR users equipped with semantics-aware premium VR devices.
	We prescribe that each VR user can only request video frames of 2D FOVs pre-cached on the MEC server side based on captured viewpoints, which is to decrease heavy data traffic load on wireless links compared to requesting 3D FOVs~\cite{sun2019communications}.
	In addition, notice that different tiles of the VR video are usually processed independently, whereas different video frames of the same tile are processed dependently~\cite{guo2021power}.
	That way, only consecutive video frames within one 2D FOV tile need to be considered at the coding level in WiserVR.
	
	In parallel, observing that static and dynamic objects may co-exist in the frame, and for explicit distinctions, we treat the dynamic objects whose positions change in different frames as ``behaviors'' and the remaining static objects as ``environments''.
	Particularly, according to their different SLGs, we devise two core sub-processes of environment reconstruction and behaviour recognition for semantic coding to align with the URLLC goal.
	In the following, we showcase the specific structure of its encoder and decoder, respectively.
	
	\subsection{WiserVR-Encoder Network}
	The WiserVR-Encoder consists of three different modules, including semantic segmentation module, SLG construction module, and channel encoder module, as illustrated in Fig.~\ref{Encoder}.
	Concretely, the first module is to segment and categorize all objects in each input video frame by employing the semantic segmentation technique, which function can be realized by deep convolutional networks~\cite{cao2022observation}.
	After that, a segmentation map and semantic labels of all objects within the frame can be obtained, respectively, where the segmentation map is a type of high-level image representation with category color label assigned to each underlying object, and each semantic label indicates a sequence of word tokens for class labels or natural language descriptions.
	With these as inputs, the second SLG construction module is able to precisely construct the SLG for each frame, which is a graph containing multiple nodes and edges with corresponding semantic labels attached.
	To be more specific, each node represents the central point of the segmentation map of each object, and then attaching respective semantic label to form the SLG of each frame.
	Note that the concept of SLG is different to knowledge graph~\cite{9416312}, where our SLG focuses more on objects' location information (e.g., the \textit{location} of an object ``a man'' in the tile, referring to the input video frame in Fig.~\ref{Encoder}), their location relationships (e.g., ``a man'' is \textit{directly below} ``a tree''), and their semantic labels (e.g., \textit{``a man'' with short blond hair in a blue sweater jacket is sitting on the ground and looking down at a book in his hands}) to provide accurate location and semantic calibration for subsequent video recovery.
	Accordingly, after comparing the SLG of each object between different frames, both environments and behaviors within each tile can be easily identified in this module.
	Moreover, semantic features (i.e., the SLG and segmentation map) of each environment are apparently identical in all video frames, which can be generalized as one frame.
	As for these behaviors, only a few frames' semantic features need to be transmitted, thanks to the powerful behavior recognition module in the subsequent WiserVR-Decoder.
	Further, in order to adapt various physical channel states (e.g., fading, interference, and signal-to-noise ratio), a channel encoder module composed with dense neural network layers is exploited to ensure these environmental and behavioral features to be accurately transmitted through different channels.
		
	\subsection{WiserVR-Decoder Network}
	\begin{figure*}[ht]
		\centering
		\includegraphics[width=1\textwidth]{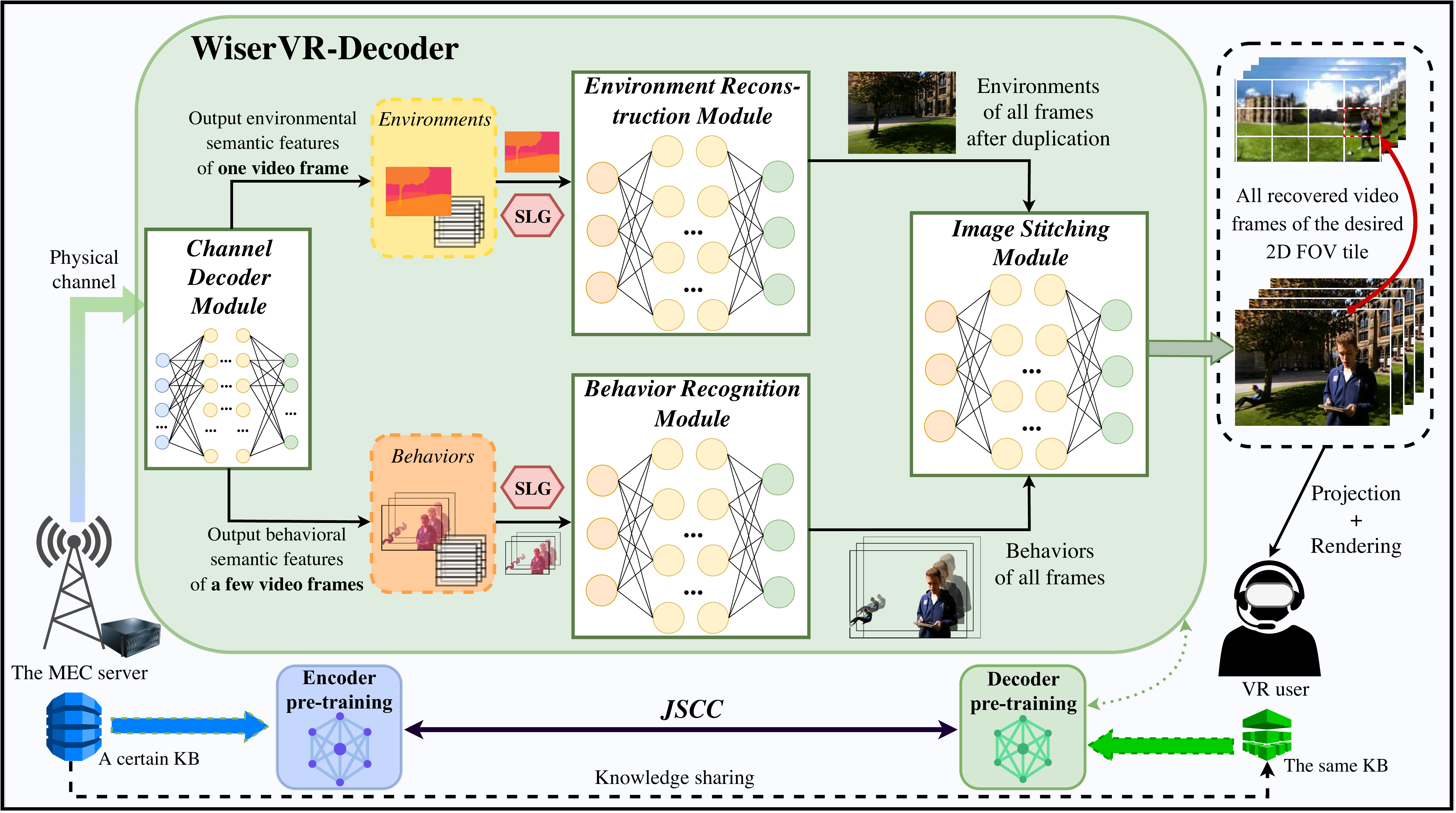} 
		\caption{The detailed structure of the WiserVR-Decoder network.}
		\label{Decoder}
    \end{figure*}
	In the WiserVR-Decoder, as demonstrated in Fig.~\ref{Decoder}, a channel decoder module with the symmetric structure of channel encoder module first recovers environmental and behavioral features within each desired 2D FOV tile, respectively.
	Based on the JSCC design, the channel decoder is capable of greatly preserving semantic features from received data stream.
	Afterward, the environmental and behavioral features will be input into an environment reconstruction module and a behavior recognition module, respectively.
	In the environment reconstruction module, the deep convolutional network~\cite{cao2022observation} can still be leveraged to rebuild all static scenes of one frame under image synthesis with multiple visual control, while the behavior recognition function is easily achieved by a Transformer model to predict all missing behaviors of those untransmitted frames~\cite{lu2022video}.
	Notably, in the both modules above, features related to segmentation map of each frame are specifically to roughly reconstruct the background image and object profiles, while the SLG is the calibration cornerstone for determining the exact location and status of each object in the tile.
	Finally, the reconstructed environments are fed into an image stitching module along with the restored behaviors to be merged together, thus obtaining all consecutive original VR video frames.
	
	\subsection{URLLC Realization with WiserVR}
	As outlined above, multiple coding models are executed in the WiserVR framework.
	In line with the URLLC requirement of VR delivery, we need to discuss the potential benefits it yields and possible promotion on latency and reliability.
	
	\textbf{Low-Latency Guaranty:}
	 In the WiserVR framework, the computing latency burden is deemed trivial at the transceiver side, since all JSCC-related coding neural networks should be well pre-trained before being put into use, which means that only little computation latency is needed for VR interaction without any further time-consuming training process.
	 Although introducing a slight extra time cost in running WiserVR, there should be a more considerable latency reduction in its communicating phase.
	 Compared to the traditional VR delivery, a large number of bits can be saved in the first place since only core semantic features along with a few video frames need to be transmitted owing to these state-of-the-art DL algorithms.
	 Apart from that, semantic features of environments and behaviors are separated by leveraging the SLG concept, thus only one frame relative to environments and a few frames relative to behaviors are encoded for VR delivery, further alleviating the heavy burden on data traffic and communication latency.
	 Comprehensively, a low latency can be assured by the proposed WiserVR.
	
	\textbf{Substantially Improved Reliability:}
	A unique feature of SemCom is the introduced background knowledge, which can provide semantic coding models with adequate semantic interpretation capabilities.
	Combined with the knowledge sharing mechanism, equivalent knowledge can be achieved at the transmitter and receiver sides, thereby ensuring highly accurate semantic recovery.
	Moreover, the JSCC approach is also applied in the WiserVR to improve its resilience and robustness against various channel states, especially for the case with severe signal impairment.
	
	\section{Implementation of WiserVR}
	\begin{figure*}[ht]
		\centering
		\includegraphics[width=0.85\textwidth]{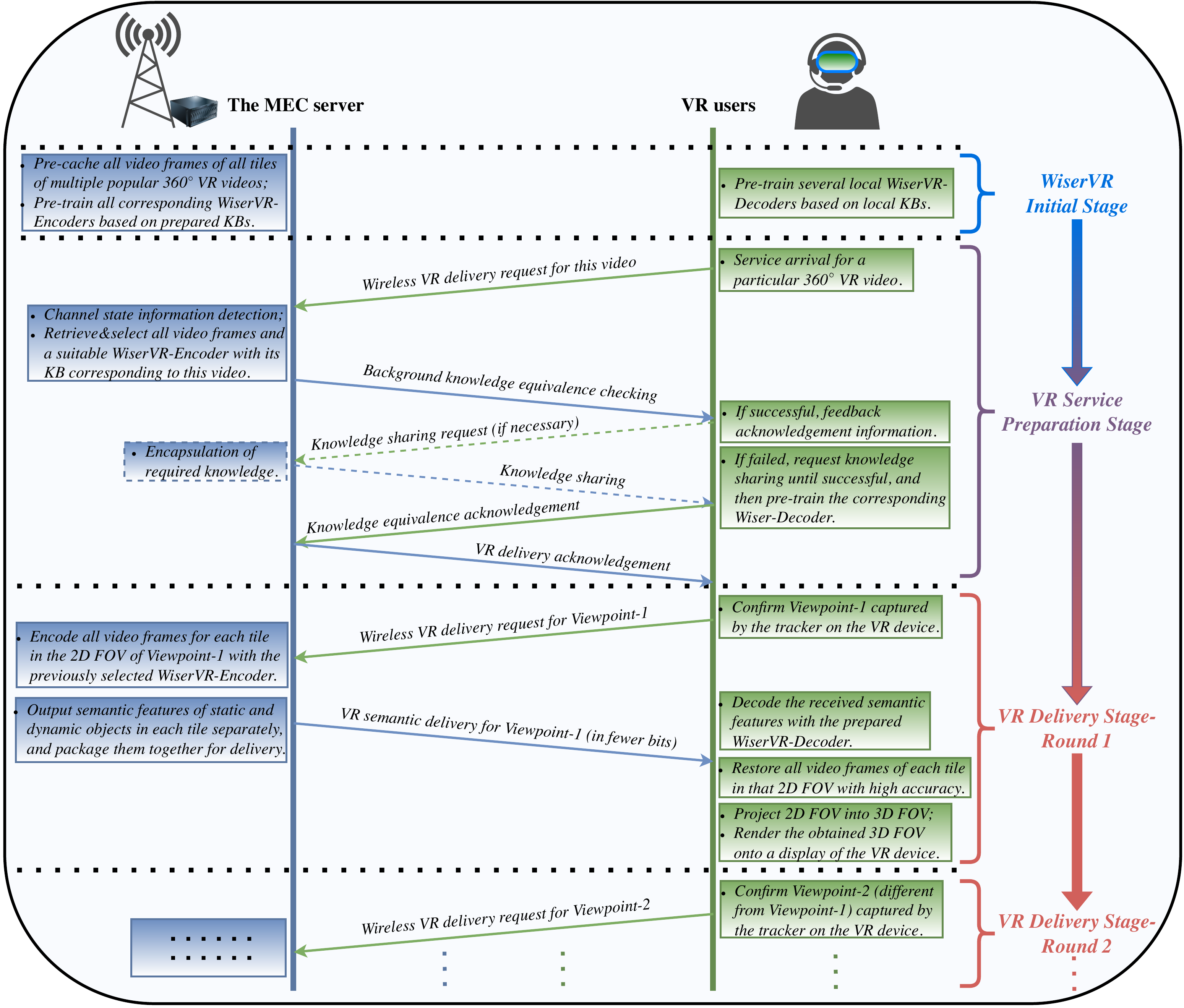} 
		\caption{A schematic diagram of implementing a particular 360$^{\circ}$ VR video delivery based on WiserVR between an MEC server and VR users.}
		\label{Implementation}
    \end{figure*}
    To further exhibit the full picture of the proposed WiserVR framework and guide network designers to make proper changes on related transmission protocols, we revisit its key rationales and depict below a viable implementation of WiserVR between a single MEC server and VR users, which is also shown in Fig.~\ref{Implementation}.
    Note that although WiserVR is dedicated to single-link VR delivery, it is capable of serving multiple VR users once a proper resource allocation is applied to multiple links.
    
    \textbf{WiserVR Initial Stage:}
	In this stage, the MEC server proactively downloads and pre-caches all same-duration video frames of all given-size tiles related to multiple popular 360$^{\circ}$ VR videos, where all video frames of each viewpoint have already finalized necessary pre-processing procedures at the cloud server, such as stitching, equirectangular projection, and FOV extraction.
	Besides, all WiserVR-Encoders that are applicable for these video frames are well pre-trained based on prepared KBs to guarantee availability on later VR delivery service requests.
	Likewise, each VR user is capable of leveraging its local storage capacity to pre-cache several local KBs and pre-train associated WiserVR-Decoders.
	
	\textbf{VR Service Preparation Stage:}
	Once the MEC server receives the service request for a particular 360$^{\circ}$ video streaming from the VR user, it detects the current wireless channel state and analyzes the service request signaling select an appropriate WiserVR-Encoder comprehensively.
	Considering the requirement of knowledge equivalence in SemCom, knowledge sharing between the MEC server and the VR user is required if the condition is triggered, and then pre-train the corresponding WiserDecoder in this case for VR semantic interpretation.
	Note here that all the delays (e.g., knowledge sharing and pre-training) occurred during the WiserVR initial stage and the VR service preparation stage are not counted into the VR delivery latency budget, which is because the first viewpoint request of the VR user has not yet started and will not affect the practical QoE.
	
	\textbf{VR Delivery Stage:}
	When it comes to this stage, both the MEC server and the VR user are basically ready for SemCom.
	Starting from the user side, there is a Viewpoint-1 of this 360$^{\circ}$ VR video captured by the tracker mounted on the mobile VR device, and correspondingly, a VR delivery request for Viewpoint-1 is sent to the MEC server.
	Afterward, all video frames of each tile in the 2D FOV of Viewpoint-1 are chosen and input together into the previously selected WiserVR-Encoder for semantic feature extraction.
	In the light of technical processes presented in the previous section, all semantic features of static and dynamic objects in each tile can be extracted separately, followed by a necessary data encapsulation procedure to package them for semantic delivery.
	Herein, it is worth re-emphasizing that much fewer bits are used for transmission in this stage when compared to the traditional one, thanks to the efficient semantic representation capability of WiserVR-Encoder.
	Moreover, based on strong restoration capabilities, the corresponding WiserVR-Decoder at the user side is capable of recovering all the video frames in each tile from received bits with high accuracy, and then combines all tiles of each timestamp into an intact 2D FOV in the spatial domain.
	Finally, the obtained 2D FOV can be projected and rendered as usual to display a crisp 3D FOV on the VR device, indicating the completion of Viewpoint-1's semantic delivery task.
	Thereafter, as the VR user may change its viewpoint (e.g., Viewpoint-2) at the next moment, a new round of VR delivery stage can be recommenced following the same process as the Viewpoint-1 task above, repeating this until the entire 360$^{\circ}$ VR video service is completely over.
	
	\section{Simulation Results and Analysis}
	In this section, simulation results are presented to evaluate the initial performance of the proposed WiserVR framework.
	For the simulation settings, an advanced deep convolutional network named Observation-Centric Sort is leveraged in the semantic segmentation module, which keeps consistent with the setup given in~\cite{cao2022observation}.
	Besides, the parameters in JSCC-related channel encoding and decoding networks are proceeding as those in~\cite{yang2022deep}, where the wireless channel model is simulated as an additive white Gaussian noise channel with the SNR varying between $-9$ dB and $6$ dB.
	Moreover, a Transformer-powered video frame interpolation (VFI) network is exploited for behavior recognition at the WiserVR-Decoder side, and its architecture details can refer to~\cite{lu2022video}.
	Finally, the Adam optimizer is adopted to train WiserVR with an initial learning rate of $5\times 10^{-4}$ based on a given video dataset.
	In parallel, for comparison purposes, we utilize two different benchmarks in the simulations: 1) A deep joint source-channel coding plus VFI (DeepJSCC) scheme~\cite{yang2022deep}, which employs a single deep neural network to delivery VR video frames and a VFI network for behavior recognition without any awareness of implicit semantics (i.e., SLG); 2) A conventional bit-oriented communication (Conventional) scheme~\cite{cai2006efficient}, in which all pixels of each video frame should be encoded into bits based on the prescribed coding rule for precise VR delivery.
	
	\begin{figure}[t]
		\centering
		\includegraphics[width=0.48\textwidth]{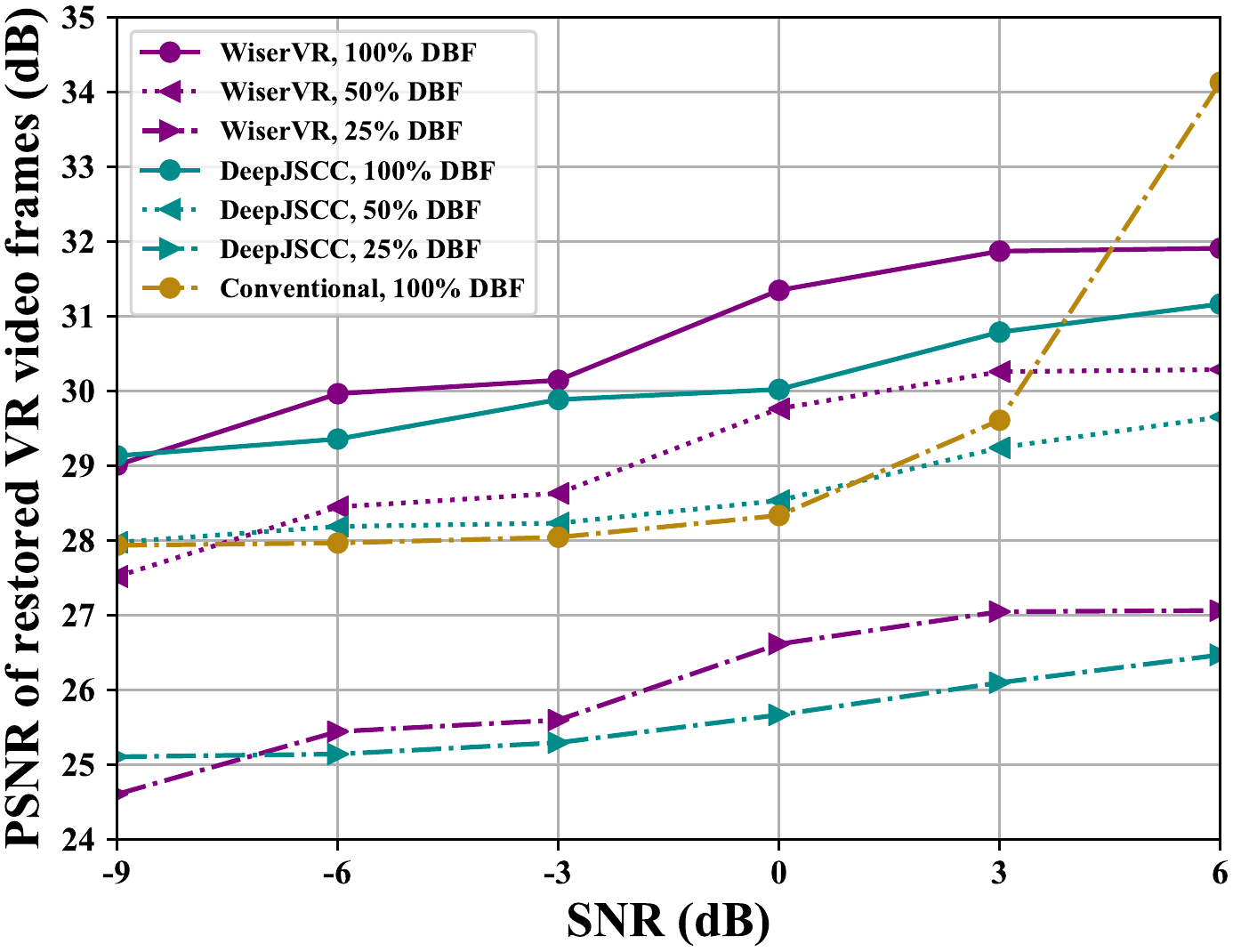} 
		\caption{PSNR of restored VR video frames versus varying SNR values.}
		\label{PSNR}
    \end{figure}
    Figure~\ref{PSNR} first verifies the frame quality of the restored VR video with different SNRs by introducing a common metric called peak signal-to-noise ratio (PSNR)~\cite{lu2022video,yang2022deep}, and generally, the higher the PSNR, the better the recovered frame quality.
    Besides, note that only a few VR video frames containing behavioral objects need to be transmitted in WiserVR, in which case we compare three different proportions about the number of delivered behavioral frames (DBF) to the number of total behavioral frames.
    It is observed that the PSNR of all schemes increases with SNR, which is because the higher SNR leads to less impairment of transmitted semantic features so as to render a more accurate frame recovery.
    Meanwhile, we can see a better PSNR of WiserVR at a higher proportion of DBF.
    This trend is attributed to the fact that using fewer behavioral frames for transmission means that more compressed features could be lost between consecutive behavioral objects, thereby resulting in a worse PSNR performance.
    Furthermore, it can be found that in the same $100\%$ DBF condition, the PSNR of WiserVR can always outperform the conventional scheme when SNR is lower than $3$ dB, while reaching an average PSNR performance gain of $0.6$ dB under all SNRs compared with the DeepJSCC benchmark.
    Such the performance superiority of WiserVR can be credited to its accurate semantic calibration function provided by SLG during video restoration, which sufficiently guarantees high reliability of wireless VR delivery even in the low-SNR regions.
    
    \begin{table}[!t]
		\centering
		\caption{Amount of Bits Required and Time Saved in Transmitting a VR Video Clip ($20$ Frames in Total)}
		\label{Time}
		\setlength{\tabcolsep}{3pt}
		\renewcommand\arraystretch{1.4}
		\begin{tabular}{|m{3cm}<{\centering}|m{2.5 cm}<{\centering}|m{2.5 cm}<{\centering}|}\hline
			\textbf{Different wireless VR delivery schemes} & \textbf{Total amount of required bits} & \textbf{Transmission time saved in percentage}\\ \hline
			WiserVR, $100\%$ DBF & $1.81\times 10^{6}$ & $93.6\%$\\ \hline
			WiserVR, $50\%$ DBF & $9.05\times 10^{5}$ & $96.8\%$\\ \hline
			WiserVR, $25\%$ DBF & $4.53\times 10^{5}$ & $98.4\%$\\ \hline
			DeepJSCC, $100\%$ DBF & $9.42\times 10^{6}$ & $66.7\%$\\ \hline
			DeepJSCC, $50\%$ DBF & $4.71\times 10^{6}$ & $83.4\%$\\ \hline
			DeepJSCC, $25\%$ DBF & $2.36\times 10^{6}$ & $91.7\%$\\ \hline
			Conventional scheme & $2.83\times 10^{7}$ & $\slash$\\ \hline		
		\end{tabular}
	\end{table}
    
    In addition, Table~\ref{Time} presents the amount of transmission bits required of all the schemes as well as the corresponding transmission time reduction compared with the conventional scheme.
    Notably, all results in this table are based on testing a transmitted VR video clip ($20$ frames in total), where the SNR of $0$ dB and the bandwidth of $1$ MHz are assumed in the transmission link between MEC server and the VR user.
    It is seen that the proposed WiserVR only requires $1.81$ Mbits under the $100\%$ DBF condition, which reduces $7.61$ Mbits compared with DeepJSCC and $26.49$ Mbits with the conventional scheme.
    In other words, WiserVR saves as high as $93.6\%$ transmission time compared with the conventional scheme under the same link quality, while DeepJSCC can only save $66.7\%$.
    Moreover, the lower proportion of DBF saves more communication resources and transmission time, which is easily understandable due to the significant reduction of delivered frames in WiserVR.
	
	\section{Open Research Issues and Outlook}
	Despite many ascendancies, our WiserVR framework still implies several inevitable and thorny issues that should be highlighted and tackled before unleashing its full potential.
	
	\textbf{Universal Semantic-Relevant VR Delivery Metrics:}
	Recall that the core of SemCom is to ensure the successful meaning delivery rather than bit transmission, which means these classical metrics, such as bit-error rate and symbol-error rate, are no longer applicable to measure its actual performance.
	Especially for SemCom-enabled wireless VR delivery, the correct reception of genuine semantics hidden behind delivered consecutive videos should be the paramount criterion when defining a universal and concrete semantic-relevant metric.
	Herein, a fresh concept of semantic base~\cite{zhang2022toward}, similar to the important role of bit in classical information theory, deserves to be further investigated in order to evolve an effective measurement for representing multimodal and multi-perspective characteristics of VR semantic information.
	
	\textbf{Heavy Semantic Coding Burden on VR Devices:}
	To achieve the optimal performance, sophisticated neural network modules need to be deployed at the mobile VR devices, imposing a huge burden on limited hardware resources (storage, memory, computational units, and battery power).
	To make WiserVR practically implementable, advanced neural network compression and acceleration technologies, including network pruning, quantization, and knowledge distillation, are feasible methods to significantly reduce the complexity of neural networks at an affordable cost of performance degradation.
	
	\textbf{VR Semantic Representation Over Time-Varying Channels:}
	Although the JSCC method has been employed to realize highly-accurate SemCom under the given different channel attributes, it is still rather challenging for us to build a semantic model to cope with the time-varying channel in practice.
	Therefore, a transfer learning-based multi-objective WiserVR framework, which incorporates pre-trained network parameters with respect to multiple common channel state information, appears to be a promising endeavor.
	
	\section{Conclusions}
	This article explored the potential of applying SemCom into VR services, for which we proposed the WiserVR framework for consecutive 2D FOV video frame delivery between the MEC server and VR users.
	Multiple DL-based modules integrated with SLG were well-devised in both the WiserVR-Encoder and WiserVR-Decoder networks, which can not only realize efficient semantic extraction and recovery, but also significantly reduce transmission traffic as well as communication latency.
	Moreover, implementation along with initial simulations are provided, followed by associated open issues and corresponding solutions.
	We hope that our WiserVR serves as a pioneer in improving communication resource usage and computational efficiency for futuristic semantics-empowered massive wireless VR delivery networks.
	
	\bibliographystyle{IEEEtran}
	\bibliography{main}
	\vspace{-10pt}
	\begin{IEEEbiographynophoto}{Le Xia} (l.xia.2@research.gla.ac.uk)
	is currently pursuing his Ph.D degree with the James Watt School of Engineering, University of Glasgow, UK. His research interests include semantic communications, intelligent vehicular networks, and resource management in next-generation wireless networks.
	\end{IEEEbiographynophoto}
	\vspace{-10pt}
	\begin{IEEEbiographynophoto}{Yao Sun} (Yao.Sun@glasgow.ac.uk)
	is currently a Lecturer with the James Watt School of Engineering, the University of Glasgow, UK. His research interests include semantic communications, intelligent wireless networking, and wireless blockchain system.
	\end{IEEEbiographynophoto}
	\vspace{-10pt}
	\begin{IEEEbiographynophoto}{Chengsi Liang} (2357875l@student.gla.ac.uk)
	is currently pursuing her Ph.D. degree with the James Watt School of Engineering, University of Glasgow, UK. Her research interest includes semantic communication and networking.
	\end{IEEEbiographynophoto}
	\vspace{-10pt}
	\begin{IEEEbiographynophoto}{Daquan Feng} (fdquan@szu.edu.cn)
	is an associate professor with the Guangdong-Hong Kong Joint Laboratory for Big Data Imaging and Communication, Shenzhen University, China. His research interests include URLLC communications, MEC, and massive IoT networks. He is an Associate Editor of IEEE Communications Letters.
	\end{IEEEbiographynophoto}
	\vspace{-10pt}
	\begin{IEEEbiographynophoto}{Runze Cheng} (r.cheng.2@research.gla.ac.uk)
	is currently pursuing his Ph.D. degree with the James Watt School of Engineering, University of Glasgow, UK. His research interests include next generation mobile networks, machine learning, blockchain system, and resource management in wireless communication.
	\end{IEEEbiographynophoto}
	\vspace{-10pt}
	\begin{IEEEbiographynophoto}{Yang Yang} (yangyang\_2018@bupt.edu.cn)
	is currently pursuing her Ph.D. degree with Beijing University of Posts and Telecommunications, China. Her research interests include device to device communications, wireless VR multimedia delivery, MEC, and distributed machine learning.
	\end{IEEEbiographynophoto}
	\vspace{-10pt}
	\begin{IEEEbiographynophoto}{Muhammad Ali Imran} (Muhammad.Imran@glasgow.ac.uk) 
	is a Professor of communication systems with the University of Glasgow, UK, and a Dean with Glasgow College UESTC. He is also an Affiliate Professor with the University of Oklahoma, USA, and a Visiting Professor at University of Surrey, UK. He has over 20 years of combined academic and industry experience with several leading roles in multi-million pounds funded projects.	\end{IEEEbiographynophoto}
\end{document}